\documentclass{ws-ijgmmp}

\begin{document}

\markboth{Dario Zappal\`a}
{Four-Dimensional Isotropic Lifshitz Points}

%
\catchline{}{}{}{}{}
%

\title{ ISOTROPIC  LIFSHITZ SCALING IN FOUR DIMENSIONS
}

\author{DARIO  ZAPPALA'
}

\address{INFN, Sezione di Catania, Via Santa Sofia 64,\\ 
Catania, 95123, Italy\\
\email{dario.zappala@ct.infn.it
} }

\maketitle

\begin{history}
\end{history}

\begin{abstract}
The presence of isotropic Lifshitz points for a O(N)-symmetric scalar theory 
is investigated with the help of the Functional Renormalization Group. 
In particular, at the supposed lower critical dimension d=4,  
evidence for a continuous line of fixed points is found for the O(2) theory, 
and the observed structure  presents  clear similarities with the properties 
observed in the 2-dimensional Berezinskii-Kosterlitz-Thouless phase.
\end{abstract}


\section{Introduction}	

Lifshitz points were first introduced in \cite{Horn}, by a generalization of the usual Landau-Ginzburg $\phi^4$  model where the space coordinates 
are treated  anisotropically  in such a way that the usual kinetic term with the square gradient of the field in one set of coordinates is  kept finite, while the square 
gradient related to the other set of coordinates is suppressed, thus promoting the term with  four powers of the gradient of the field to the leading kinetic role.
This class of points also includes the case of isotropic Lifshitz points where, for all coordinates, the four gradient term is leading, the square gradient term 
being suppressed.

These points are associated to the class of tricritical points on the phase diagram, where ordered and disordered phase coexist with a third phase
where the order parameter shows a periodic structure with finite wave vector. It has several realisations in condensed matter,
such as magnetic systems, but also polymer mixtures, liquid crystals, high-Tc superconductors (for reviews on this subject see \cite{erzan,sak,diehl}), 
and also  in the formulation of emergent gravity \cite{horava,filippo,cognola}  as well as 
 in the analysis of dense quark matter with the realization of unconventional phases \cite{casal2,nardulli,buballa,pisarski}.

In this paper, after reviewing  some general properties of the  Lifshitz points in Section 2,  we discuss in Section 3  the isotropic points  for $O(N)$ 
theories, which show up  in the  dimensional range  $4 < d < 8$, by means of a non-perturbative approach, namely the Functional Renormalization 
Group (FRG) flow equations \cite{Wetterich:1992yh, Morris:1994ie, Berges:2000ew}, which is especially useful in cases that are out  
of reach of other approaches such as the $\epsilon$-expansion \cite{Horn,erzan,diehl4}. 
Then, in Section 4, we concentrate on the case $d=4$ \cite{zappa, zappaprd}, where we observe very interesting properties associated to the Lifshitz scaling,
such  as the appearance of a continuous line of fixed points,  that present clear similarities with those observed in the $d=2$
Berezinskii - Kosterlitz -Thouless phase \cite{berezinskii71, kosterlitz73}.
Our conclusions are reported in Section 5.

\section{General Properties of Lifshitz Points}

The general form of the action $S[\phi]$, suitable for describing a Lifshitz point in $d$ dimension, with $m$-dimensional anisotropic scaling, is \cite{Horn}:
\begin{equation}
S=\int {\rm d}^{D}x_{_\bot }{\rm d}^{m}x_{_\parallel} \left [ \frac{W_{_\parallel} }{2} (\partial_{_\parallel} ^2 \phi)^2  
+  \frac{W_{_\bot }}{2} (\partial_{_\bot }^2 \phi)^2  + \frac{ Z_{_\parallel} } {2}  (\partial_{_\parallel}  \phi)^2 
+ \frac{  Z_{_\bot }}{2} (\partial_{_\bot } \phi)^2 +V \right ]
\label{startaction}
\end{equation}
where $D=d-m$, and $\phi(x)$ is a $N$-component vector field  and the potential $V= V(\phi)$ is a generic function of the field.
Equation~(\ref{startaction}) allows for  an anisotropic structure with different scaling properties  of the two subsets of coordinates,
$x_\parallel$  (which is $m$-dimensional)   and   $x_\bot$ ,  (which is $(d-m)$-dimensional) that is realized by taking 
$Z_\parallel = 0$ and $Z_\bot=1$. Then, in the orthogonal directions the leading derivative term is the 
one with two  derivatives of the field and the term proportional to  $W_\bot$ with four derivatives remains irrelevant,  
while the absence of a two derivative term in the parallel directions makes the term with four derivatives, proportional to $W_\parallel$,
the leading kinetic term in this subset of coordinates.

 In addition, at the mean field level, a negative  $Z_\parallel < 0$, with  values of the square mass in $V(\phi$) below a critical value,
 induces  the appearance of a new modulated phase  with an oscillating ground state and, in particular, the 
 point $Z_\parallel =m^2=0$ corresponds to a tricritical point, indicating  the coexistence  of this phase with the two other phases:
 disordered ($<\phi>=0$) and ordered (constant $<\phi>\neq 0$). 
  
The presence of two different kinetic terms leads to two different scaling regimes with scales $\kappa_\bot$ and  $\kappa_\parallel$
of  the two subsets of coordinates, for instance in the two-point  function of the theory, and therefore to two different anomalous dimensions, 
$\eta_{ l 2}$ and  $\eta_{l 4}$. These, in turn  connect the the scales $\kappa_\bot$ and  $\kappa_\parallel$ through  the  anisotropy parameter 
$\theta=(2- \eta_{ l 2})\,/\,({4 -\eta_{ l 4} })$, according to the relation $\kappa_\parallel=\kappa_\bot^\theta$. \\
Moreover, this twofold  scaling implies the following scaling dimension of the field \\ $d_\phi =({  d-m +\theta (m-4+ \eta_{ l 4} )})/ { 2}$
in units of $\kappa_\bot$ and other operators appearing in  Eq.~(\ref{startaction}), transform  accordingly.

\begin{figure}[ph]
\centerline{\psfig{file=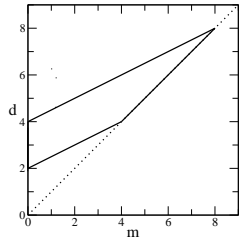,width=1.9in}}
\vspace*{8pt}
\caption{Quadrilateral region of convergence of the integral in Eq.~(\ref{integrale}), in the $(m,d)$ plane.
\label{figu1}}
\end{figure}

Another peculiar aspect of the Lifshitz point is given by the region of the $(m,d)$ plane where quantum corrections do not show singular behavior.
In order to have a basic  indication on this region, one can determine the  intervals of $d$ and $m$ where the one loop integral that contributes 
to the  two point function  ( $ D=d-m$)
\begin{equation}
I_{d,m}({\rm p,q})= \int \frac{{\rm d}^m  {\rm q' }}{(2\pi)^m}\frac{{\rm d}^{D}{ \rm p'}}{(2\pi)^{D}} \; \frac{1}{ \rm p'^2+q'^4}  \; \frac{1}{| {\bf p+ p' } |^2+  | {\bf q+ q' } |^4 } 
\label{integrale}
\end{equation}
does not show any pathology  either in the infrared or in the ultraviolet region \cite{parisshpot}. 
One  easily realizes that the relevant region corresponds to the quadrilateral 
displayed in Fig.~\ref{figu1}, delimited by a segment of the straight line $d=4+m/2$ on the upper and one of the 
line $d=2+m/2$ on the lower side, plus the segment $2<d<4$ on the $d$ axis on the left, and a segment on the line $d=m$ on the right side.

Therefore, Fig.~\ref{figu1} shows the lowest order indications for the upper and lower critical dimensions of the anisotropic Lifshitz point extracted 
by a simple  dimensional analysis and which depend both on $d$ and $m$.
Clearly this picture is  modified by higher order corrections and in particular by the effective change in the dimensions induced by a non-vanishing 
anomalous dimension that has been neglected so far.

We now focus on the specific case of interest, namely the isotropic Lifshitz point, defined by $m=d$, i.e. the parallel subspace coincides with the full 
$d$-dimensional space and no orthogonal subspace is left :
\begin{equation}
S=\int {\rm d}^{d}x_{_\parallel} \left [ \frac{W_{_\parallel} }{2} (\partial_{_\parallel} ^2 \phi)^2  
+ \frac{ Z_{_\parallel} } {2}  (\partial_{_\parallel}  \phi)^2  +V \right ] \,.
\label{isoaction}
\end{equation}

In this case the rotational  symmetry in the $d$ dimensional  space is fully recovered but, unlike the 
standard case where the scaling  is set by the two derivative term, in Eq.~(\ref{isoaction}) 
the scaling is  determined by the four derivative  term.
In fact all coordinates now scale as  $\kappa^{-1}_\parallel$ and the anomalous dimension associated with 
$\kappa_\bot$ is vanishing, $\eta_{ l 2}=0$, so that the anisotropy parameter reduces to $\theta=2\,/\,({4 -\eta})$, where $\eta\equiv \eta_{ l 4} $.

Therefore, it is more convenient to define the scaling dimensions of the various operators with respect to $\kappa_\parallel$ and then  the scaling 
dimension of the field is $d_\phi =({ (d-4+ \eta)})/ { 2}$, while the dimensions of $V$, $W_\parallel$, $Z_\parallel$, are respectively 
$d$, $-\eta$, $2-\eta$. Accordingly, a square mass operator $m^2$ and a quartic coupling $u$ appearing in the potential $V(\phi)$,
have dimensions respectively $4-\eta$ and $8-d -2\eta$. 

In addition, we remark that the values of the upper and lower critical dimensions, corresponding respectively to a vanishing scaling dimension of $u$: 
$d_u=8-2\eta$ and to a vanishing scaling dimension of $\phi$ : $d_l=4-\eta$, contain the corrections, due to the anomalous dimension, to the lowest order 
indications coming from Fig.~\ref{figu1}  with $d=m$, where one finds $d_u=8$ and $d_l=4$. We also find an additional relevant operator, with positive 
scaling dimension, namely $Z_\parallel$, that has no counterpart in the usual picture of the standard dimensional analysis.

\section{Functional Renormalization Group Analysis}

We shall now briefly summarize some results  concerning  the determination of the Lifshitz point and in particular of the corresponding
anomalous dimension,  obtained with the help of the Functional Renormalization Group flow equations, that can be regarded as an alternative 
non-perturbative tool to complement the known perturbative techniques such as the $\epsilon$-expansion (for the Lifshitz  case,  $\epsilon= 8-d$)
\cite{Horn,erzan,diehl4},
or the $1/N$-expansion (for a $O(N)$ theory with large number of fields) \cite{horn2}.
The  functional differential equation that determines the FRG flow is \cite{Berges:2000ew}:
\begin{equation}
\label{rgfloweq}
k \partial_k  \Gamma_k[\phi]=\frac{1}{2} \int_q\, \partial_t R_k(q)
\left [ \Gamma_k^{(2)}[q,-q;\phi]+R_k(q)\right ]^{-1}
\end{equation}
$\Gamma_k[\phi]$ being the running  effective action at scale $k$,
$\Gamma_k^{(2)}[q,-q;\phi] $ its second functional derivative with respect to the field,
and  $R_k(q)$  a suitable regulator that suppresses the
modes with $q\ll  k$ and allows to integrate those with  $q\gg k$.
The flow runs from an ultraviolet fixed action, that is taken as a boundary condition of the differential 
equation at $k=\Lambda$, down to  the deep infrared region, $k=0$, where 
the solution of the differential equation, $\Gamma_{k=0}[\phi]$ results in the full effective action,
i.e. the 1PI diagram generator of the theory.  The fixed points of the theory appear as stationary 
solutions of the flow equation, properly expressed in terms of dimensionless quantities.

In order to solve the flow equation, one has to introduce some approximation scheme, as well as some
specific ansatz of the running effective action (as, for instance, the explicit form given in Eq.~(\ref{isoaction}),
with $k$-dependent parameters  $W_{_\parallel }, \,Z_{_\parallel }, \, V$). 
Then, the full flow reduces to a set of flow equations for  the $k$-dependent parameters.
\begin{figure}[ph]
\centerline{\psfig{file=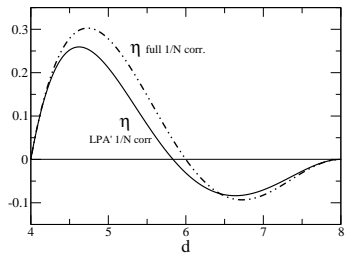,width=2.1in}}
\vspace*{8pt}
\caption{The anomalous dimension $\eta$ as obtained in the large $N$ expansion at order $1/N$ (dot-dashed)
and in the LPA' approximation (solid).
\label{figu2}}
\end{figure}

The case of a scalar $O(N)$ symmetric theory, was considered long ago in the framework of the $1/N$-expansion
and in particular the $1/N$ order computation of the anomalous dimension of the isotropic Lifshitz point 
as a function of the dimension $d$, was carried out \cite{horn2,gubser}. The result is reported in Fig.~\ref{figu2}.

In addition, in  Fig.~\ref{figu2}, it is shown the output of the FRG determination of the anomalous dimension at order $1/N$ ,  obtained in 
\cite{zappa} by first determining the flow equations for the various parameters to order $1/N$ and then  by solving these equations 
at two different orders of approximation, namely the Local Potential Approximation (LPA) (where  $W_{_\parallel }=1, \,Z_{_\parallel }=0, \eta=0$
are kept fixed and only one flow equation for the potential  $V_k(\phi)$ is solved), and  the improved approximation LPA' (where 
again $W_{_\parallel }=1, \,Z_{_\parallel }=0$ and their flow is neglected but, together with the flow of the potential, one allows for a non-vanishing 
anomalous dimension $\eta$).

The results of the LPA indicate the existence of a nontrivial Lifshitz point when $4 < d < 8$, as expected from the analysis discussed 
in the previous Section. The result of the LPA'  for $\eta$ are shown in Fig.~\ref{figu2}. As discussed above, $\eta$ from the LPA' is not the
complete $1/N$ determination of the anomalous dimension, as it comes from the LPA' approximation to the flow equation at order $1/N$. 
In spite of that, the agreement of the two determinations in Fig.~\ref{figu2} is qualitatively good and they become coincident when approaching
$\eta=0$ in proximity of $d=4$ and $d=8$ \cite{zappa}.

In conclusion,  the expectations of a Lifshitz point in the range $4 < d < 8$  (as $\eta$ vanishes at the two end points) at large $N$
are  confirmed both by the full $O(1/N)$ computation and by the LPA' approximation to the $O(1/N)$ FRG flow equations. 
Remarkably, $\eta$ switches sign from positive to negative,  with a zero  around $d=6$.

The picture in the case of the  single field scalar theory ($N=1$), is instead less clear, at least in proximity of $d=4$. In fact  a perturbative approach 
in this case consist in an $\epsilon$-expansion around the upper critical dimension $d=8$, and therefore a non-perturbative tool
is needed at the other endpoint. Here, we just briefly mention the FRG numerical  analysis of the coupled equations for 
$W_{_\parallel }, \,Z_{_\parallel }, \, V$  (see \cite{boza}), in the context of the Proper Time Flow, as in this case the differential equation for these three 
parameters had already been derived in \cite{litimzappala}. The output of this analysis is that a non-trivial Lifshitz point with negative anomalous dimension
is found in the range $5.5 < d < 8$. Above $d=8$, as expected, only a gaussian-like  Lifshitz point  exists, while the lower limit, $d=5.5$, does not have
a physical meaning, as it is only a numerical limit below which the solution becomes very difficult to find, probably because of interference with multicritical
solutions  that appear  when $d< 5.5$.

\section{Isotropic Lifshitz points in $d=4$}

The results of the previous Sections show the nice correspondence between the range $ 2 < d < 4$ for the standard scaling and the range
 $ 4 < d < 8$ for the isotropic Lifshitz scaling, due to the modification in the scaling dimension of the field. For the latter 
 scaling regime, at least for large  $N$, the presence of a non-trivial Lifshitz point, which disappears both for $d > 8$ and for $d < 4$, is verified
 in analogy with the Wilson-Fisher fixed point, observed in the former scaling regime with $ 2 < d < 4$.   
 A relevant  difference between the two cases is the change of sign of the Lifshitz anomalous dimension, whereas  $\eta>0$ 
in  the Wilson-Fisher case.

By following the above analogy, one could expect that the Lifshitz scaling shows, at the lower critical dimension $d=4$, 
properties that are similar to those observed at the lower critical dimension of the standard scaling, $d=2$. 
In particular in  $d=2$, it is known that  despite the  Coleman - Mermin - Wagner theorem forbids, for a $O(N)$ theory, 
an ordered phase with finite order parameter  and, therefore, also forbids a typical phase transition from 
a disordered to an ordered phase,
it is still possible to observe, for the $O(2)$ theory, a transition of topological nature from a disordered to a 
quasi-ordered phase (i.e. with algebraic, rather than exponential, decay of the two-point correlation function at large distance), which is known as 
Berezinskii - Kosterlitz - Thouless  (BKT) transition\cite{berezinskii71, kosterlitz73}.

Although for the Lifshitz scaling in $d=4$ there is no equivalent of the 
Coleman-  Mermin - Wagner theorem, we now show that some scaling features
observed in the case of the BKT transition, are also reproduced in the Lifshitz case.
In fact,  a few interesting results are obtained by means of a simple FRG analysis, 
performed in analogy with a previous study on the two-dimensional BKT transition \cite{jame}.

Our starting point is the following four dimensional  $U(1)$ invariant model :
\begin{eqnarray} \label{toymbis}
&\Gamma_k [\phi] &  =\int d^4{\bf r}  \Bigg \{ \frac{u_k}{8} \, \left( |\phi |^2 - \alpha^2_k \right)^2 +
\frac{W^A_k}{2} \left [ \partial^2 \phi \, \partial^2  \phi^* \right ] 
\nonumber \\ 
&& + \frac{W^B_k}{8} \,  \left [ \partial^2 |\phi |^2 \right]^2  + \frac{Z^A_k}{2} \left [ \partial \phi \, \partial  \phi^* \right ] +  \frac{Z^B_k}{8}  \,  
\left [ \partial |\phi |^2 \right]^2
\Bigg\}
\end{eqnarray}
where four field derivative as well as  two field derivative terms, with field independent parameters, $W^A_k,\, W^B_k, \, Z^A_k, \, Z^B_k$,
have been included and the quartic  potential is expressed in terms of  $u_k, \, \alpha_k$.
Instead of a $O(2)$ symmetric action, in  Eq.~(\ref{toymbis}) we took an action invariant under $U(1)$ transformations of
the complex field $\phi({\bf r})$, that can be  decomposed into a longitudinal and a transverse component, including an 
expectation value of the longitudinal component  $\alpha_k$:  $\phi({\bf r}) = \alpha+ \sigma({\bf r}) + i\pi({\bf r}) $. 

In this scheme one can compute the flow equation for the various $k$-dependent parameters \cite{zappaprd}.  
By starting the flow at an initial scale $k=\Lambda$,
with large values of $\alpha_\Lambda$, one immediately observes for  this parameter a power law  when $k\to 0 $, with exponent $\eta$, 
according to its scaling dimension : $\alpha^2_k \propto k^\eta$.  At the same time the field renormalization parameter $W^A_k$ shows the 
inverse scaling $W^A_k\propto k^{-\eta}$, again in agreement with the scaling dimensional analysis. As a consequence, the renormalized square field,
i.e. the product $J=W^A_k a_k^2$,  remains constant along the flow.   
When the flow is started at a lower value  of $\alpha_\Lambda$, the above picture breaks down and $J$  shows a scale  dependence that leads
to $J\to 0$ at some finite scale $k$.

\begin{figure}[ph]
\centerline{\psfig{file=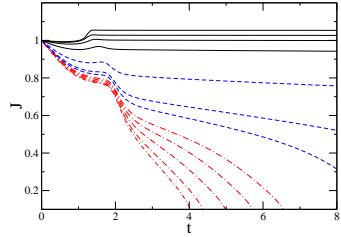,width=2.0in}}
\vspace*{8pt}
\caption{ Flow of $J=W^A_k a_k^2$ vs. $t=\log(\Lambda/k)$. for different initial values of $\alpha_\Lambda$. Flat upper solid curves correspond to larger 
values of $\alpha_\Lambda$,  while dashed and dot-dashed to lower values.
\label{figu3}}
\end{figure}

This  behavior is summarized in Fig.~\ref{figu3} ,where $J$ is plotted vs. $t=\log(\Lambda/k)$ for different initial values of $\alpha_\Lambda$,
with upper (solid) curves corresponding to larger initial values. Each  upper curve in Fig.~\ref{figu3} indicates
the presence of a fixed point where all dimensionless parameters in our model reach a $k$-independent value and remain constant along the flow,
 as verified in \cite{zappaprd}. In particular,  the relevant parameter $Z^A_k$, when its initial value is suitably taken on the critical surface, in 
 the infrared region goes to zero as $Z^A_k\propto k^{2-\eta}$ .
Therefore, we obtained a continuous line of fixed points  parametrized by $J$ (or equivalently by $\alpha_\Lambda$)
 that disappears  at sufficiently small $J$, as the corresponding curves (dashed in Fig.~\ref{figu3}) are no longer constant. 
 
 This is  analogous to the picture obtained for the 
 BKT  transition where $\alpha^2_k \to 0 $, when $k\to 0$, indicates a vanishing order parameter both in the disordered 
 and in the quasi-ordered phase, while the line of fixed points associated to a large non-vanishing  $t$-independent $J$ 
 (in BKT language, the stiffness) 
 indicates the presence of a quasi-ordered phase associated to the algebraic decay of the two-point function.

It is known that the BKT line of fixed points, regardless of
the microscopic details of the model considered, 
ends at the universal value $J=2/\pi$ and,  below it, 
$J$ is no more scale independent, as $J\to 0$, and
the disordered phase replace the quasi-ordered phase. 
Apparently, this endpoint corresponds,  in  Fig.~\ref{figu3},  
to the separation between the upper solid flat curves  and  the lower
dashed curves that  show no constant behavior. 

Unfortunately, this statement is not exactly true.  In fact, there is no clear 
 transition between the two sets of curves and, more specifically, 
if one continues the flow of the upper set to extremely large  values of $t$, one will eventually observe a decrease of $J$ toward zero, 
meaning that in Fig.~\ref{figu3}, strictly speaking, there is no
definite value of $J$ signalling a transition and the regime corresponding to the line of fixed points is only approximately reached at large $J$.

This, in turn, does not mean that for the present case of the Lifshitz point, there is no equivalent of the BKT transition, but only that the 
approximations involved in the FRG analysis could be not sufficient to detect this effect, yielding only an approximate description of it.
In fact even for the two dimensional BKT case, the analogous FRG analysis produces similar results 
\cite{jame, grater},  and  attempts to reproduce the  correct picture require an enlargement of the ansatz in Eq.~(\ref{toymbis}) and 
therefore a  greater numerical effort \cite{dupuis}, or the use of  more sophisticated approaches \cite{defenu, krieg},  
not  applied so far to the more complicated case of the Lifshitz scaling.
 
At this point, instead of testing  improved versions of the FRG flow, we look for a possible field configuration that could explain
the scaling properties  discussed above. To this purpose,  we consider   Eq.~(\ref{toymbis}) as the starting point, and use  
polar coordinates $\phi({\bf r})=\sqrt{ \rho ({\bf r}) }\, {\rm exp }[i\theta ({\bf r})]$.  Then, in the infrared region 
a mass term for the radial component  $\rho({\bf r})$ suppresses  spatial fluctuations, so that this component reduces to a constant
$\rho({\bf r}) \to \rho_0$. In addition, we neglect relevant operators such as the two field derivative term, that are suppressed 
in the infrared region, when suitably taken on the critical surface. Therefore, we consider the 
following effective action depending  on the angular fluctuations only,
\begin{equation}
\Gamma=\frac{K}{2} \int d^4{\bf r}  \left [ \partial^2 \theta({\bf r})  \, \partial^2  \theta({\bf r})  \right ] \; ,
\label{hamil} \end{equation} 
to provide the correct description of the infrared sector of the original  theory. \\
In principle,  Eq.~(\ref{toymbis}) produces an additional term proportional to 
$(\partial \theta  \, \partial  \theta)^2$ that we neglect here, as it is always possible to  include in Eq.~(\ref{toymbis})
an additional operator, proportional to $\left [ \partial \phi \, \partial  \phi^* \right ]^2$ , that cancels exactly the quartic  term in $\theta$.
  
Then, by recalling the form of the Green function of the Lapacian operator in four dimensions, 
$\partial_{\bf r}^2  \; [-1/({\bf r} - {\bf r'})^2 ] = (2\pi)^2 \delta^4 ({\bf r} - {\bf r'})$, one realizes that  the configuration
\begin{equation}
\theta_{\bf r'}  ({\bf r}) =   \int d^4{\bf r''}  \frac{1}{(2\pi)^2}  \frac{1}{({\bf r} - {\bf r''})^2 } \; \frac{1}{({\bf r''} - {\bf r'})^2} =\frac{1}{4} \;
{\rm  ln} \left (\frac{R^2}{ ({\bf r} - {\bf r'})^2  } \right )
\label{conf} 
\end{equation} 
which has a  singularity  at the point ${\bf r'}$, does actually minimize $\Gamma$  in Eq.~(\ref{hamil}), as the extremum equation gives: 
$( \partial_{\bf r}^2\,  \partial_{\bf r}^2)  \theta_{\bf r'}  ({\bf r}) = (2\pi)^2 \delta^4 ({\bf r} - {\bf r'})$, 
i.e. it vanishes  everywhere, but at the singular point ${\bf r'}$.
The integral in Eq.~(\ref{conf}) is then performed by introducing a large distance cut-off $R$; 
the output is displayed in the right hand side of~(\ref{conf}). 

This result resembles the one  obtained for the vortex configuration of  the BKT problem and, as in that case,  it allows to compute 
the energy associated to the configuration in Eq.~(\ref{conf}), provided one introduces also a short distance cut-off (or lattice spacing), $r_0$.
In fact, by regarding Eq.~(\ref{hamil}) as the hamiltonian of the angular field $\theta ({\bf r})$ in a  4-dimensional space,
one gets the energy for the configuration in Eq.~(\ref{conf}):  $\Gamma [ \theta_{\bf r'}  ] =\frac{K}{2} \pi^2 \, {\rm  ln} \left ({R^2 / r^2_0  } \right )$.
Then, by noticing that the entropy $\Sigma$ associated to placing such a configuration (i.e. the singularity ${\bf r'}$) in the four dimensional space delimited
at large and small distance respectively  by $R$ and $r_0$, is given by   $ \Sigma[ \theta_{\bf r'}  ] ={\rm  ln} \left ({R^4 / r^4_0  } \right )$, one can estimate 
the free energy $F$ of the system  ($T$ indicates the temperature) :
\begin{equation}
F =\Gamma[ \theta_{\bf r'}  ] - T \, \Sigma[ \theta_{\bf r'}  ] = \left (  \frac{K}{2}  \pi^2  - 2\,T \right )\, {\rm  ln} \left ({R^2 / r^2_0  } \right )
\label{free} 
\end{equation} 
Then, as for the BKT case, the transition of $F$ from positive to negative indicates instability with respect to the generation of such configurations 
and therefore indicates the transition to a disordered phase where free (unpaired) configurations are observed . The transition point  in our case is
$K/T= 4/\pi^2$, and although this specific value  directly depends on the particular normalization chosen in (\ref{conf}), it signals the existence of a
transition point in $K/T$, from the quasi-ordered to the disordered phase, that was not detected in the previous FRG analysis.

 \section{Conclusions}
 
The FRG analysis of the Isotropic Lifshitz scaling shows, more accurately in the case of very large number of fields $N$,
the presence of a non-trivial fixed point in the dimensional range $4 < d < 8$ that, to some extent, resembles the properties 
of the  well known Wilson-Fisher fixed point in the range  $2 < d < 4$. Exactly as for the latter case in  $d=2$, it is found that the
Lifshitz point disappears in $d=4$. 

However, it has been found, again
in close analogy with the two-dimensional BKT transition,  that for the $O(2)$ symmetric model in $d=4$, the Lifshitz scaling 
predicts the presence of a continuous line of fixed points, corresponding to a quasi-ordered phase with algebraic 
long-distance decay of the correlation function.

Unfortunately, due to the limits of the approximations adopted to solve the FRG flow,
no definite picture of a phase transition to the disordered phase is obtained. 
 Instead of attempting to  improve on these approximations, we conjectured that a specific configuration of the
 angular component of the complex field,  which is a  minimum of  the energy, is responsible of the 
 fixed point line and also of the transition to the disordered phase  at some finite value of the effective coupling,
 in the same way as the vortex configurations act in the BKT transition.

\section*{Acknowledgments}

This work has been carried out within the Istituto Nazionale di Fisica Nucleare (INFN) project QFT- HEP.

\end{document}